\documentclass[superscriptaddress,
reprint,longbibliography,
preprintnumbers,nofootinbib,
amsmath,amssymb,aps,prl]{revtex4-1}

\pdfoutput=1

\usepackage[utf8]{inputenc} 
\usepackage[T1]{fontenc} 
\usepackage{microtype}
\usepackage{mathrsfs}
\usepackage{amsmath}
\usepackage{physics}
\usepackage{balance}
\usepackage{natbib}
\usepackage{adjustbox}

\usepackage{graphicx}
\makeatletter
\def\@biblabel#1{[\textbf{#1}]}
\makeatother
\usepackage[dvipsnames]{xcolor}

\usepackage{dcolumn}
\usepackage{bm}

\usepackage[colorlinks = true,
            linkcolor = blue,
            urlcolor  = blue,
            citecolor =red,
            anchorcolor = blue]{hyperref}
\usepackage{verbatim}
\usepackage{color,ulem}
\usepackage[english]{babel}

\usepackage[scr]{rsfso}

\usepackage{blindtext}

\begin{document}

\preprint{IFT-UAM/CSIC-20-55}

\title{\Large Physics of Phonon-Polaritons in Amorphous Materials}

\author{Luigi Casella}%
 \email{luigi.casella@studenti.unimi.it}
\affiliation{
Department of Physics "A. Pontremoli", University of Milan, via Celoria 16, 20133 Milan, Italy.
}

\author{Matteo Baggioli}%
 \email{matteo.baggioli@uam.es}
\affiliation{Instituto de Fisica Teorica UAM/CSIC, c/Nicolas Cabrera 13-15,
Universidad Autonoma de Madrid, Cantoblanco, 28049 Madrid, Spain.
}%

\author{Tatsuya Mori}
\email{mori@ims.tsukuba.ac.jp }
\affiliation{Division of Materials Science, University of Tsukuba, 1-1-1 Tennodai, Tsukuba, Ibaraki 305-
8573, Japan}

\author{Alessio Zaccone}%
 \email{alessio.zaccone@unimi.it }
\affiliation{
Department of Physics "A. Pontremoli", University of Milan, via Celoria 16, 20133 Milan, Italy.\\
Department of Chemical Engineering and Biotechnology,
University of Cambridge, Philippa Fawcett Drive, CB30AS Cambridge, U.K.\\
Cavendish Laboratory, University of Cambridge, JJ Thomson
Avenue, CB30HE Cambridge, U.K.
}

\begin{abstract}
The nature of bosonic excitations in disordered materials has remained elusive due to the difficulties in defining key concepts such as quasi-particles in the presence of disorder. We report on the experimental observation of phonon-polaritons in glasses, including a boson peak (BP), i.e. excess of THz modes over the Debye law. A theoretical framework based on the concept of diffusons is developed to model the broadening linewidth of the polariton due to disorder-induced scattering. It is shown that the scaling of the BP frequency with the diffusion constant of the linewidth strongly correlates with that of the Ioffe-Regel (IR) crossover frequency of the polariton. This result demonstrates the universality of the BP in the low-energy spectra of collective bosonic-like excitations in glasses, well beyond the traditional case of acoustic phonons, and establishes the IR crossover as the fundamental physical mechanism behind the BP.

\end {abstract}

\maketitle
\textit{Introduction}.
The low-energy vibrational spectra of solids provide direct insights into the complex many-body atomic dynamics of materials~\cite{Born-Huang}. Understanding the vibrational spectra is crucial for our understanding and technological design of the optical, thermal and mechanical  properties of solids.
Substantial experimental and theoretical efforts have focused on the case of phonons in amorphous materials, where phonons are well-defined quasi-particles only in the limit of long wavelengths. On shorter length-scales, disorder dominates the vibrational excitations and gives rise to deviations from Debye's quadratic law in the vibrational density of states (VDOS), resulting in the boson peak in the Debye-normalized VDOS detected originally in Raman scattering spectra~\cite{Shuker}. The origin of the boson peak in phonon spectra remains controversial.
A line of research has traditionally supported the identification of the BP with shifted and smeared van Hove (VH) singularities~\cite{Elliott,Monaco}. However, recent studies have pointed out that the boson peak may be largely independent, and in fact even unaffected by the lowered VH singularity. This is indicated by the co-existence of the BP with the lowest (transverse) VH singularity in the spectra of simple models systems~\cite{Milkus,Walter_vanHove,PhysRevResearch.2.013267}.

Another line of research points at the close link between BP and the Ioffe-Regel (IR) crossover between ballistic phonons and quasi-localized excitations~\cite{PhysRevLett.96.045502,Shintani2008,PhysRevB.87.134203}, as the origin of the BP. Among the theoretical frameworks, the most popular is the heterogeneous elasticity theory of Schirmacher, Ruocco and co-workers~\cite{Ruocco2007,Schirmacher,Mizuno} based on the assumption of spatial correlations in the shear elastic modulus, in agreement with simulations~\cite{Mizuno}.

\begin{figure}[t!]
    \centering
    \includegraphics[width=0.9\linewidth]{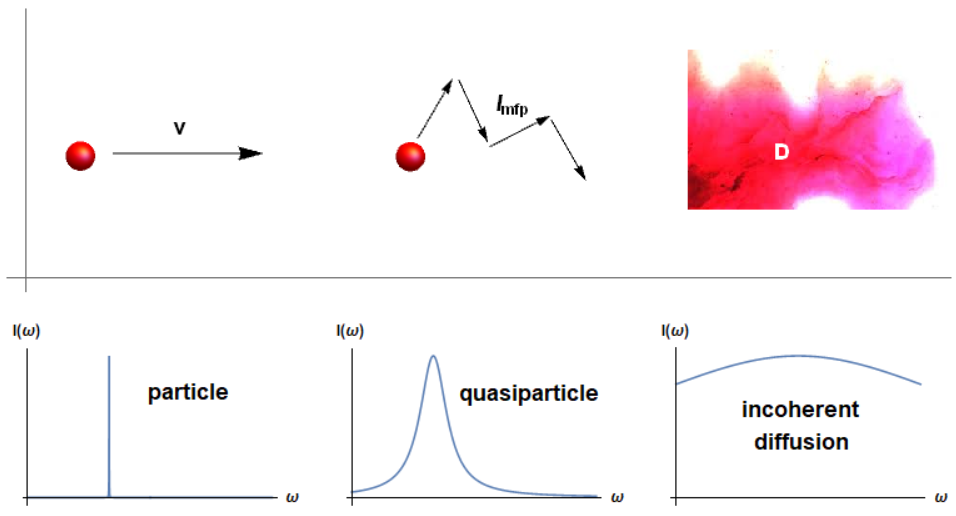}
    \caption{The destruction of the quasiparticles induced by disorder-scattering. The dynamics of the system becomes totally incoherent and collective and its low frequency dynamics is well described by hydrodynamics.}
    \label{example}
\end{figure}

Although growing consensus is emerging about the crucial role of \textit{randomness} in driving the crossover from ballistic to quasi-localized excitations leading to a BP, a picture supported by random matrix theory~\cite{Zaccone_review,Schirmacher_random,PhysRevE.100.062131}, the physical origin of the BP remain controversial. Furthermore it remains to be established whether the BP is a distinctive feature of the phonon spectra or if it is a more general phenomenon common to all bosonic-like excitations in amorphous solids (e.g. excitons, plasmons, other polaritons). 

Here we provide an answer to these fundamental questions by reporting on the experimental observation of the BP in infrared absorption spectra of phonon-polaritons in a model glass, i.e. soda-lime silicates. A theory of phonon-polaritons in amorphous materials is presented, which clarifies the origin of the BP from the Ioffe-Regel (IR) crossover between the ballistic quasi-particle (coherent) propagation to a quasi-localized regime dominated by disorder-induced scattering, where the quasi-particle loses its coherence and undergoes diffusive-like dynamics (\textit{diffusons})~\cite{AllenFeldman,PhysRevB.87.134203,PhysRevResearch.2.013267} (see Fig.\ref{example}).\\

These results show the that the BP is a truly universal feature for all bosonic excitations in amorphous materials, not just phonons. Also, the theoretical analysis shows that the BP is a genuine result of the Ioffe-Regel crossover caused by disorder since the polariton spectra are not affected by VH singularities. Finally, the theoretical analysis shows that the BP does not originate from the flattening of the polariton dispersion relations.

\vspace{0.2cm}

\textit{Theoretical model} -- We start by modelling the coupled dynamics of the optical phonon modes and the EM field. The first and fundamental step is the dynamical equation for the relative atomic displacement $\vec{u}$, 
\begin{equation}
\label{displacement field}
    \Ddot{\vec{u}} = -\,\omega_{0}^{2}\,\vec{u} + f(\dot{\vec{u}}) + b_{12}\,\vec{E}.
\end{equation}
The relative displacement field characterizes the relative  motion of the partially-charged particles and applies to optical (and not acoustic) vibrational modes.
The first term in the  r.h.s. of Eq.\eqref{displacement field} defines the characteristic  frequency of the harmonic oscillator, which originates from the linear restoring force acting on the atoms. The second term is an effective damping contribution, where $f$ is some function. The last term is a direct dipole coupling to the  external electric field $\vec{E}$ due to the partial charge carried by the atoms.\\
In absence of damping, $f(\dot{\vec{u}})=0$, this problem was considered in \cite{Born-Huang,Huang-1951} where  Eq.\eqref{displacement field} was solved together with an equation for the polarization that contains the effects of the relative displacement of the atoms, and with the Maxwell equations for the EM field.  Notice that damping is introduced in Eq.\eqref{displacement field} only in the  mechanical part of the equations but not in the EM sector. Upon identifying $f(\dot{\vec{u}})=\Gamma(k)\dot{\vec{u}}$, we obtain a quartic equation for the modes $\omega(k)$ which reads:
\begin{align}
\label{Eq. dispersion - damping function1}
  &  \omega^{4}\,\varepsilon_{\infty} + i \,\omega^{3}\, \Gamma\left(k\right) \varepsilon_{\infty} - \omega^{2}\left(\omega_{0}^{2}\,\varepsilon_{0} + k^{2}c^{2} \right) \nonumber\\&-i \, \Gamma\left(k\right) k^{2}c^{2} \omega + \omega_{0}^{2} \,k^{2}\,c^{2} = 0.
\end{align}
In the above equation, $c$ is the speed of light, $\omega_0$ the characteristic frequency (the energy gap of the optical mechanical mode), $\varepsilon_{0},\varepsilon_{\infty}$ the dielectric constants at zero and infinite frequency, respectively. Finally, $\Gamma(k)$ defines the linewidth, and hence lifetime $\tau$, of the optical mode, $\tau^{-1}\sim \Gamma$. The details of the derivation of Eq.\eqref{Eq. dispersion - damping function1} are  provided in ~\cite{Suppl}.

Equation~\eqref{Eq. dispersion - damping function1} can be viewed in two different ways. First, one can assume the momentum $k \in \mathbb{R}$ to be real, and the frequency to be complex. In this framework, the  modes $\omega(k)=\mathrm{Re}\,\omega(k)+\mathrm{i}\,\mathrm{Im}\,\omega(k)$ are  usually referred to as \textit{quasinormal modes} and the imaginary part of the frequency determines their exponential decay in time $\sim e^{-\mathrm{Im}\omega(k) t}$. Alternatively, one could take the frequency as real and the momentum to be complex. In this case Eq.\eqref{Eq. dispersion - damping function1} can be solved for $k(\omega)$ and the imaginary part of the momentum determines the exponential decay in space $\sim e^{-\mathrm{Im}k(\omega) x}$ -- i.e. the penetration length. The two scenarios are interchangeable. In the rest of the manuscript we will use the abbreviations $\mathrm{Re}\,k \equiv k'$ and $\mathrm{Im}\,k \equiv k''$.\\ 

By using the linear relation between the polarization vector $\vec{P}$ and the electric field $\vec{E}$, we can derive the  \textit{dielectric function},
\begin{equation}
    \varepsilon(\omega, k) = \varepsilon_{\infty} - \dfrac{\omega_{0}^{2}\left(\varepsilon_{0} -\varepsilon_{\infty}\right)}{\omega^{2}-\omega_{0}^{2} + i\, \omega \, \Gamma\left(k\right)}
\end{equation}.

Furthermore, by studying the spatial exponential attenuation (Lambert-Beer) of the wave intensity $I$,
\begin{equation}
    I / I_{0} = e^{- \alpha(\omega) x} = e^{- 2 \,k''(\omega)\, x}
\end{equation}
we can define the \textit{absorbance coefficient} $\alpha(\omega)$, which is the inverse of the \textit{penetration length}~\cite{Bornatomic}. Given a collection of scattering centers, the mean free path is given by $\ell = (\sigma n)^{-1}$, where $\sigma$ is the scattering cross-section and $n$ the number density of scatterers. It can be shown that $dI / dx = - I n \sigma = - I / \ell$~\cite{Bornatomic}, leading to exponential attenuation $I \sim \exp(-x/\ell)$. Upon comparing this with the above equation, we thus obtain 
\begin{equation}
\label{elle}
\ell^{-1} = 2 \,k''(\omega)
\end{equation}
a relation that will be useful also later on.

An  expression for the absorbance coefficient can be derived using the complex dielectric function for EM radiation in continuous media $\varepsilon(\omega)= c^2\,k^{2}/\omega^{2}$~\cite{landau_electro}. Solving this expression for $k$ and taking the imaginary part (see Eq. (4)), we find:
\begin{equation}
\label{eq:abs and dielectric}
        \alpha(\omega)= \dfrac{\omega}{c}
        \sqrt{2 \left(\, \left|\varepsilon(\omega)\right| - \mathrm{Re}\,\varepsilon(\omega)\,\right)}\,.
\end{equation}

At this point, it is crucial to specify the nature of the linewidth $\Gamma(k)$, which is neglected in standard treatments \cite{Huang-1950,Huang-1951}. The linewidth encodes the effects of the disorder on the propagation of the polariton.  The basic idea is that disorder can be represented as a large number of "defects", each acting on the polariton quasi-particle as a (elastic~\cite{Ruocco2007}) scattering center. On scales larger than the defects average separation, the result of a great number of scattering events is the diffusion of momentum through the system. This effective description is based on the idea of \textit{diffusons} \cite{AllenFeldman, Allen1993}, a concept which proved useful in explaining the anomalies in thermal transport observed experimentally in glasses \cite{ZellerPhysRevB.4.2029}. A diffusive linewidth for phonons can indeed explain the ubiquitous appearance of a boson peak in the vibrational density of states (VDOS) of glasses \cite{PhysRevLett.122.145501,PhysRevResearch.2.013267} and even the presence of a linear in $T$ term in the specific heat at low $T$ \cite{PhysRevResearch.1.012010}. Moreover, the diffusive nature of the linewidth is supported by Random Matrix Theory \cite{BeltukovPhysRevB.87.134203,PhysRevE.100.062131}. 

Following ~\cite{Shintani2008,BeltukovPhysRevB.87.134203,PhysRevResearch.2.013267}, we take the linewidth to be of diffusive form
\begin{equation}
\label{hydro}
    \Gamma(k)\,=\,D\,k^2
\end{equation}
which follows from an effective hydrodynamic treatment \cite{landau2013fluid} for quasi-particle excitations or simply from diffusion of momentum in the governing dynamic equation for the displacement field~\cite{PhysRevResearch.2.013267}. This expression is supported by experiments and simulations~\cite{Baldi2010,Shintani2008,PhysRevB.87.134203,Ruzicka,Szamel} and is valid only at relatively large $k$, while it is expected to break down at larger momenta where hydrodynamics is no longer a good approximation.\\

In Fig.\ref{fig10}, in the Supplemental material~\cite{Suppl}, we show the Debye-normalized absorbance coefficient obtained from this model for a wide range of values of the diffusion coefficient $D$. The absorbance coefficient is directly proportional, up to a linear growing function of the frequency denoted as $C(\omega)$~\cite{tat5}, to the VDOS. As a consequence, an excess in $\alpha(\omega)/\omega^2$ corresponds to a boson peak in the normalized VDOS $g(\omega)/\omega$. We observe that the BP moves to lower energies by increasing the diffusion constant $D$ and it becomes sharper. In the inset, we show the dispersion relation of the corresponding phonon-polariton modes obtained from Eq.\eqref{Eq. dispersion - damping function1} and we compare it with the BP frequency $\omega_{BP}$. This dynamics and the underlying physics mechanism behind will be discussed in detail later.

\vspace{0.2cm}

\textit{Comparison with experimental data} -- The linewidth ceases to display a hydrodynamic diffusive behaviour as in Eq.\eqref{hydro} at large momenta approaching the molecular size.
We observe that the high-frequency part of the experimental spectra is well-fitted by a constant damping coefficient:
$\Gamma(k)\,=\,\gamma$,
which corresponds to a Langevin friction term in Eq.\eqref{Eq. dispersion - damping function1}, as expected in local molecular-level dynamics in glassy environment. Indeed, on small length-scales, we cannot coarse grain the effects of disorder into a hydrodynamic description but we have to consider the high-frequency microscopic dynamics~\cite{Zaccone_review} producing a momentum-independent relaxation time $\tau^{-1}\sim \gamma$.\\

\begin{figure}[h!]
 \centering
\includegraphics[width=0.9\linewidth]{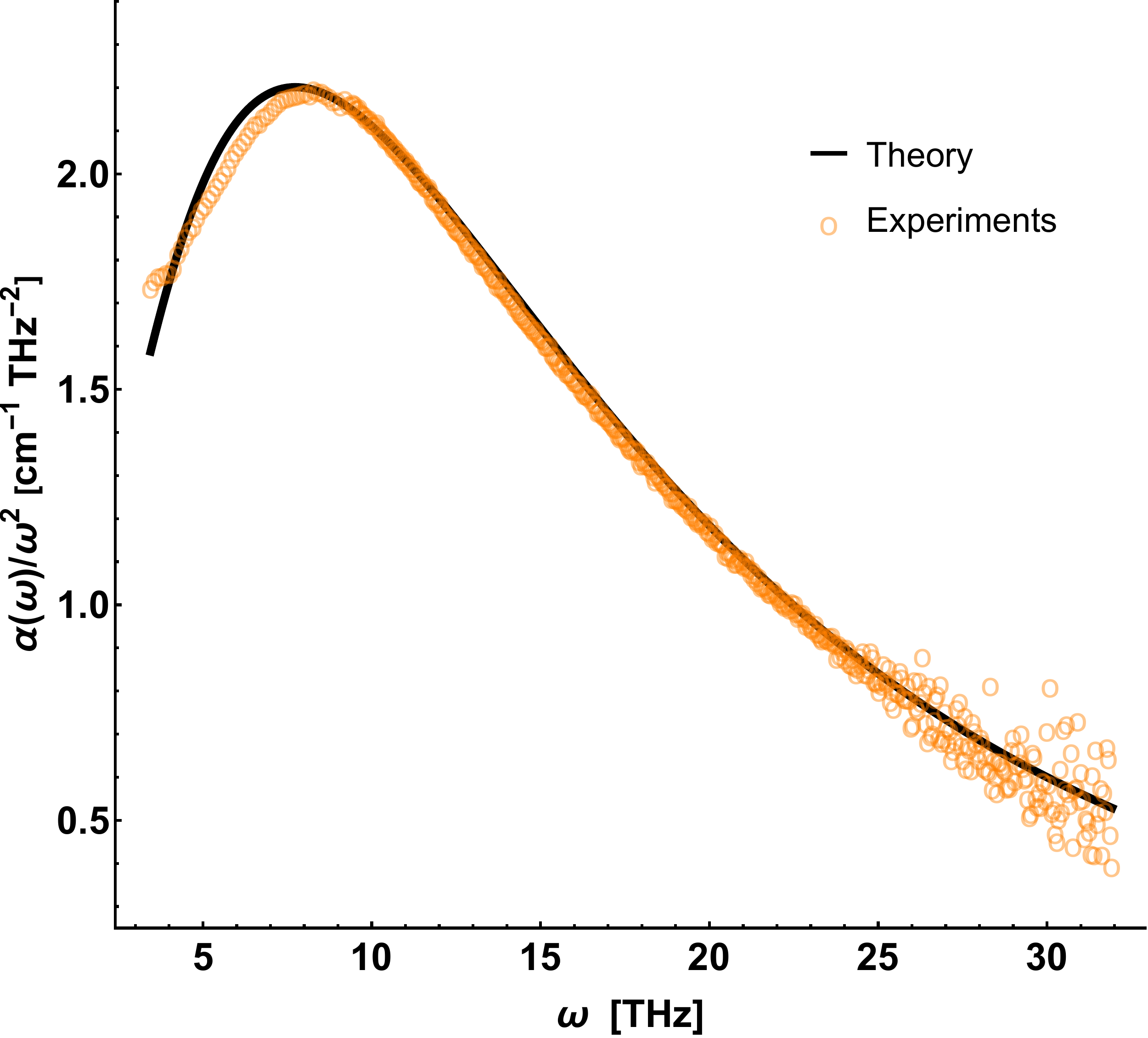}
\caption{The comparison between the theoretical model based on the diffusive linewidth function Eq.\eqref{tutto} (black line) and the experimental data for a soda lime glass (orange cirlces). The fit gives: $D=13.99 \text{THz}^{-1}/c^{2}$, $\gamma=69.67 \text{THz}^{-1}$, $\epsilon_0=70.99$, $\epsilon_\infty=61.11$, $\omega_0=29.80 \text{THz}$.}
\label{fig:intAbsor}
\end{figure}

In order to have a good description of the experimental data across the entire range of momenta, we will consider a linewidth which interpolates from the diffusive form \eqref{hydro} at low $k$ to the Langevin constant damping $\gamma$ at large $k$. More specifically, we use an interpolating model of the form:
\begin{equation}
    \Gamma(k)\,=\,\dfrac{\gamma \,D\, k^{2}}{\gamma + D\, k^{2} }\label{tutto}
\end{equation} 
which retrieves the two limits.
We test our theoretical model using experimental measures of the infrared-absorbance spectra on a soda-lime glass sample using two different THz time-domain spectrometers that can cover a wide frequency range between 0.3 – 5 THz  (see Supplemental Material ~\cite{Suppl}).
The excellent agreement between the theoretical predictions and the experimental data is shown in Fig.\ref{fig:intAbsor}. The \textit{diffusons} behaviour at low $k$ is key to obtain a good agreement at low frequencies and we checked that it cannot be described with simple damped harmonic oscillator (DHO) model as shown in ~\cite{Suppl}. It is also to be noted that the decay at large frequencies is dominated by the constant damping term $\gamma$.

\vspace{0.2cm}

\textit{The Origin of the Boson Peak} -- Our effective theoretical model gives an accurate qualitative description of the experimental data and it is able to reproduce the boson peak. We should now address the question of the fundamental physical origin of the BP in the polariton spectra.
Let us recall that the BP frequency is defined as
\begin{equation}
    \omega_{BP}\,:\quad \quad  \frac{d }{d\omega}\dfrac{\alpha(\omega)}{\omega^{2}}\Big|_{\omega_{BP}}\,=\,0
    \end{equation}
and it corresponds to the maximum in the Debye-normalized absorbance spectra.\\

A possible explanation for the BP could come from the flattening of the phonon-polariton band:
\begin{equation}
    \omega_{flat}\,:\quad \quad  \frac{d\omega}{dk}\Big|_{\omega_{flat}}\,=\,0
\end{equation}
which, similarly to the van Hove singularities in ordered crystals, would produce a peak in the VDOS  since $g(\omega)\sim (d\omega/dk)^{-1}$~\cite{kittel2004introduction}. As one can readily verify in Fig.\ref{fig1}, the position of the BP does not correspond to the flattening of the lowest branch. Hence, the flattening of the polaritonic dispersion relations cannot satisfactorily explain the occurrence of the BP.\\

From a different perspective, it is well-known that waves in amorphous and disordered systems stop to propagate ballistically at a certain frequency known as the Ioffe-Regel frequency $\omega_{IR}$ \cite{ioffe1960non}. Moreover, the correlation between the Ioffe-Regel frequency and the BP frequency has been observed and discussed in recent works \cite{Shintani2008,PhysRevLett.96.045502,PhysRevB.87.134203}.

The Ioffe-Regel frequency is defined as the energy at which the mean free path of the wave $\ell$ becomes comparable to its wavelength $\lambda$~\cite{ioffe1960non},
 	\begin{equation}
 	\label{mfp and lambda}
    \ell(\omega_{IR} ) = \lambda(\omega_{IR})
\end {equation}
and its quasiparticle nature is lost.
Upon combining Eq.~\eqref{mfp and lambda} and Eq.~\eqref{elle}
and $k' = 2\pi/\lambda$, we obtain
\begin{align}
    &  \omega_{IR}\,:\quad \quad  k'(\omega_{IR})=4 \pi\,k''(\omega_{IR})  \label{defIR}
\end{align}
which provides a new operational quantitative definition of the Ioffe-Regel frequency for a generic collective excitation.
\begin{figure}[h!]
\centering
\includegraphics[width=0.9 \linewidth]{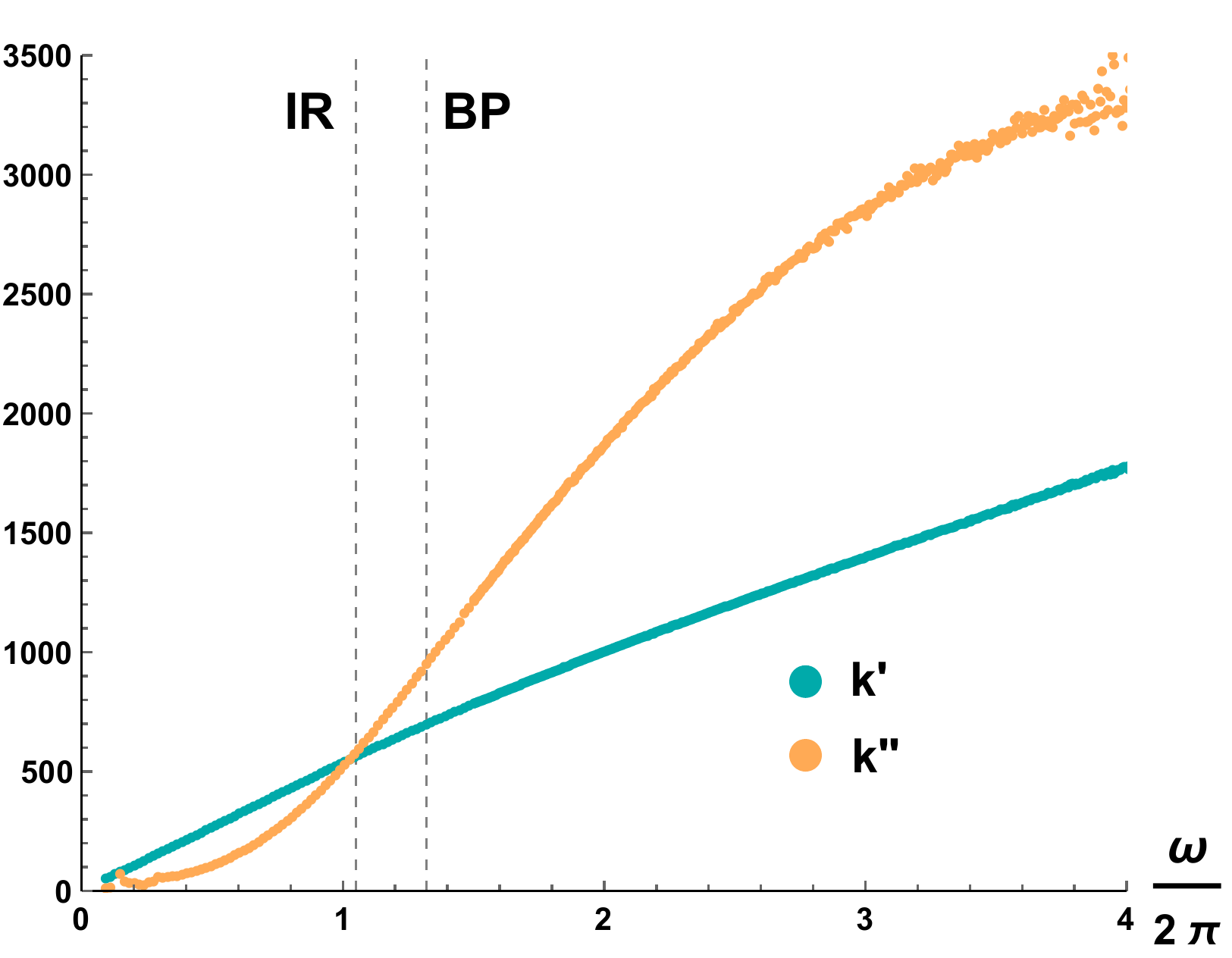}
\caption{The experimental data for $k(\omega)$ in the soda lime glass (see  fig.\ref{fig:visAbsor} in the Supplementary Material \cite{Suppl}). The first dashed line indicates the location of the IR frequency defined as \eqref{defIR}; the second dashed line indicates the BP position which can be found from the absorbance data in Fig.\ref{fig:intAbsor}. We find that $\omega_{BP}/\omega_{IR}\,\sim\,1.25$, confirming the results of Eq.\eqref{meraviglia}.}
\label{fig:BPeakIRegel0}
\end{figure}\\
In Fig.\ref{fig:BPeakIRegel0}, we show the experimental data of $k(\omega)$ for the soda lime glass (see more details in the Supplemental Material~\cite{Suppl}). From there it is evident that the Ioffe-Regel crossover, defined using Eq.\eqref{defIR}, is extremely close to the BP frequency observed in the absorbance (Fig.\ref{fig:intAbsor}), $\omega_{BP}/\omega_{IR}\,\sim\,1.25$. This represents a strong experimental confirmation of the intimate correlation between the BP and the IR frequencies in the phonon-polariton spectrum of glasses.\\

\begin{figure}[h!]
\centering
\includegraphics[width=0.9 \linewidth]{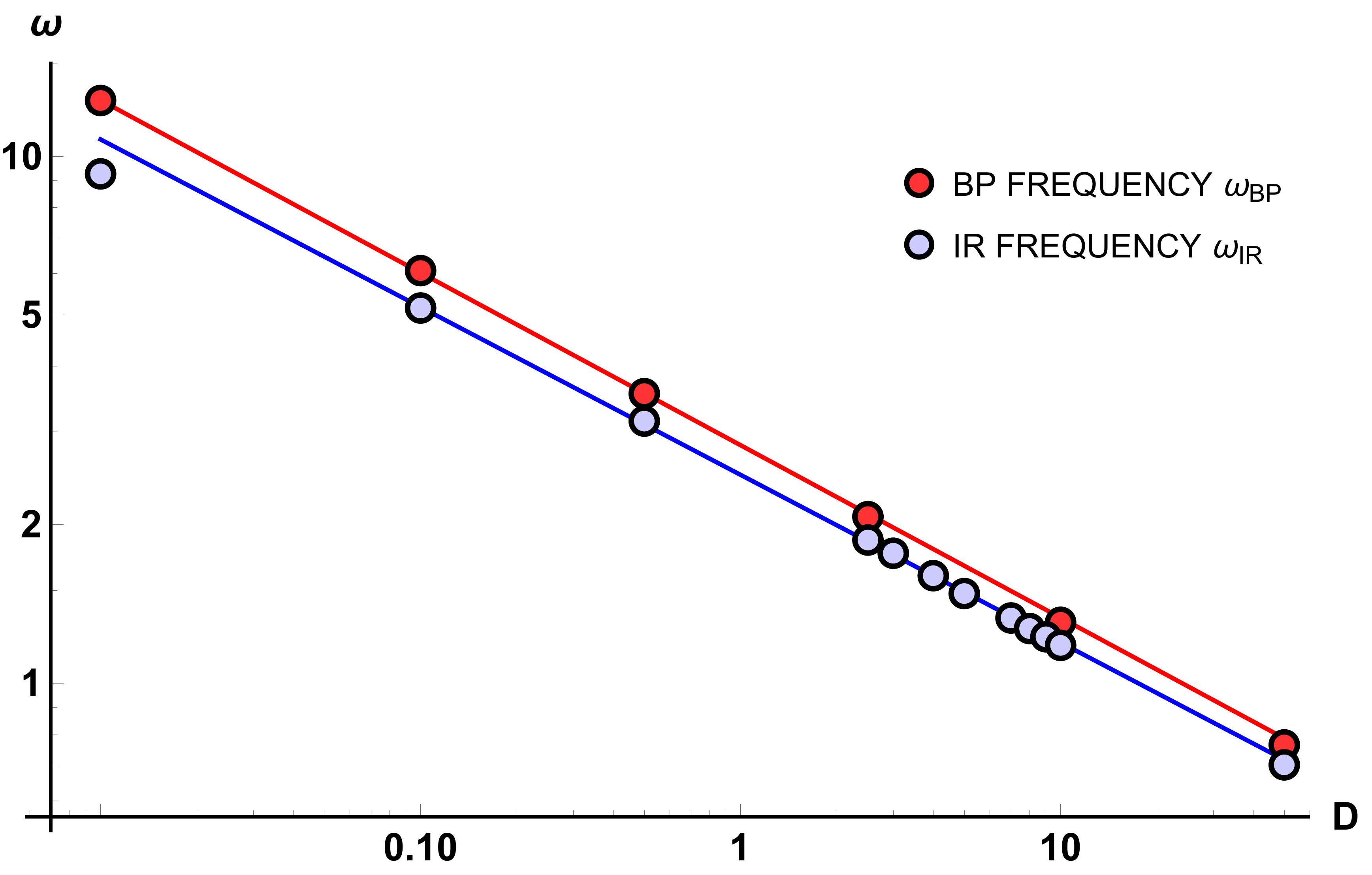}
\caption{The BP frequency $\omega_{BP}$ and the IR frequency $\omega_{IR}$ as a function of the diffusion constant $D$. The red and blue lines show the fit which at large $D$ is consistent with a scaling $\omega \sim D^{-n}$ with $n \approx 0.32$.}
\label{fig:BPeakIRegel}
\end{figure}

In order to emphasize our point, we compare the BP frequency $\omega_{BP}$ and the Ioffe-Regel frequency $\omega_{IR}$ for the theoretical model in Fig.\ref{fig:BPeakIRegel}. We again observe that the two values strongly correlate:
\begin{equation}
    \omega_{BP}\,=\,\mathcal{C}\,\omega_{IR} \label{meraviglia}
\end{equation}
where $\mathcal{C}$ is an $\mathcal{O}(1)$ constant prefactor. From the soda lime glass experimental data we consistently find $\mathcal{C}\sim\,1.25$ (see Fig.\ref{fig:BPeakIRegel0}).\\

Importantly, we also find that both frequencies follow an approximate power law scaling $\omega \sim D^{-n}$ with $n \approx 0.32$, which strengthens the idea of correlation between these two quantities. More broadly, we expect Eq.\eqref{meraviglia} to hold generally, up to a non-universal $\mathcal{O}(1)$ number -- $\mathcal{C}$ -- which depends on the microscopics of the system.

\vspace{0.2cm}

\textit{Conclusions} -- In summary, we reported on the experimental observation and theoretical analysis of phonon-polaritons in a model amorphous material.
The polaritonic nature of the excitation cannot be reproduced by standard DHO fitting, but only using a  diffusive $\sim k^{2}$ linewidth.
We can confidently claim that the boson peak observed experimentally in the phonon-polariton absorption spectrum is controlled by the Ioffe-Regel crossover from ballistic quasi-particle propagation to incoherent diffusive-like excitations (\textit{diffusons}~\cite{AllenFeldman}). This identification, which is valid with high precision, suggests that the physical mechanism underlying the BP in the phonon-polariton spectra of glasses is due to the  quasi-localization of the excitations and to the propagating-to-diffusive crossover \'a la Ioffe-Regel. Working with polaritons has the advantage that we could clearly rule out the influence of dispersion relation band flattening on the peak, away from the influence of pseudo-van Hove singularities, and hence this analysis provides the first unambiguous demonstration that boson peak and Ioffe-Regel crossover coincide, in an experimental system.
Crucially, our results suggest that this mechanism for the BP may apply to  \textit{any} bosonic excitation in amorphous materials (such as excitons, magnons, plasmons)~\cite{DasSarma}, which opens up new opportunities for technological design and control of optical, electrical and thermal properties of materials by tailoring the disorder-induced effects.\\

\vspace{0.2cm}

\textit{Aknowledgments} -- M.B. acknowledges the support of the Spanish MINECO’s ``Centro de Excelencia Severo Ochoa'' Programme under grant SEV-2012-0249. A.Z. acknowledges financial support from US Army Research Laboratory and US Army Research Office through contract nr. W911NF-19-2-0055. T.M. is grateful to Y. Matsuda for providing the glass sample and acknowledges JSPS KAKENHI Grant Nos.
JP17K14318 and JP18H04476, and the Asahi Glass Foundation.

\bibliographystyle{apsrev4-1}
\bibliography{amo}

\appendix
\section{\Large Supplementary Material}
\subsection*{Experimental Methods}
\begin{figure}
    \centering
    \includegraphics[width=\linewidth]{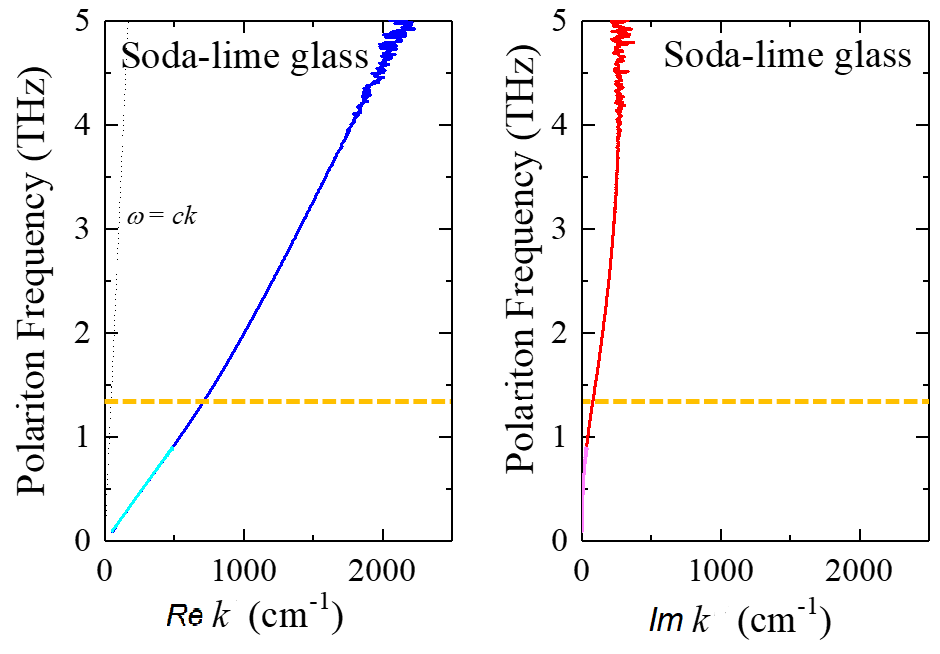}
    \caption{The dispersion relation of the Polariton in the Soda-lime glass extracted from the experimental data used in this work. The yellow dashed line indicates the BP frequency $\omega_{BP}/(2\pi)\sim 1.32 $ THz.}
    \label{didi}
\end{figure}

As a standard glass system exhibiting the BP in the infrared spectrum, we selected a soda-lime glass which is a typical network glass former. The sample is purchased from Central Glass Co., Ltd. We utilized two different commercial THz time-domain spectrometers to cover a wide frequency range between 0.3 – 5 THz (0.3 - 1.2 THz: RT-10000, Tochigi Nikon Co.; 1.2 – 5 THz: TAS7500SU, Advantest Corp.) \cite{tat1,tat2,tat3,tat4,tat6doi:10.1080/00150193.2016.1214522,tat7doi:10.1063/1.4901657}. The measured THz waveforms including multiple reflections in the sample surfaces were converted to the frequency domain, and the obtained complex transmission coefficient $t$ was analyzed using the following equation:
\begin{equation}
    t(\omega)\,=\,t_{vs}\,t_{sv}\,\frac{e^{i(n_s-1)\,d\,\omega/c}}{1-r_{sv}^2(\omega)\,e^{i\,2\,n_s\,d\,\omega/c}}\,,\label{exp1}
\end{equation}
where:
\begin{equation}
    t_{ij}\,=\,\frac{2\,n_i}{n_i\,+\,n_j}\,,\quad r_{ij}\,=\,\frac{n_i\,-\,n_j}{n_i\,+\,n_j}
\end{equation}
are the complex Fresnel’s transmission and reflection coefficients, respectively, at the interface between regions $i$ and $j$. The subscripts $i$ and $j$ stand for $v$ and $s$ in equation \eqref{exp1}, representing the vacuum and sample, respectively. $n_i$ is the complex refractive index of region $i$, $d$ is the thickness of sample, and $c$ is the speed of light. Then, the absorption coefficient $\alpha(\omega)$ is obtained from the relation:
\begin{equation}
    \alpha(\omega)\,=\,\frac{2\,\omega\,\kappa(\omega)}{c}\,,
\end{equation}
where $\kappa(\omega)$ is the imaginary part of $n_s$, i.e. the extinction coefficient. From the linear response theory for disordered systems \cite{tat5}, $\alpha(\omega)$  and the vibrational density of states $g(\omega)$ are related through the infrared photon-phonon coupling coefficient $C_{IR}(\omega)$ as following:
\begin{equation}
    \alpha(\omega)\,=\,C_{IR}(\omega)\,g(\omega)
\end{equation}
The BP appears in the spectrum of $g(\omega)/\omega^2$, therefore the BP in the infrared spectrum appears in the plot of $\alpha(\omega)/\omega^2$.\\
Some experimental data for the dielectric constant are shown in Fig.\ref{fig:visAbsor} together with the fits from the theoretical model. Moreover, in Fig.\ref{didi} we show the dispersion relation of the polariton as extracted from the experimental data.
\section{Theoretical model calibration on experimental data}
\begin{figure}[t]
 \centering
\includegraphics[width=0.9\linewidth]{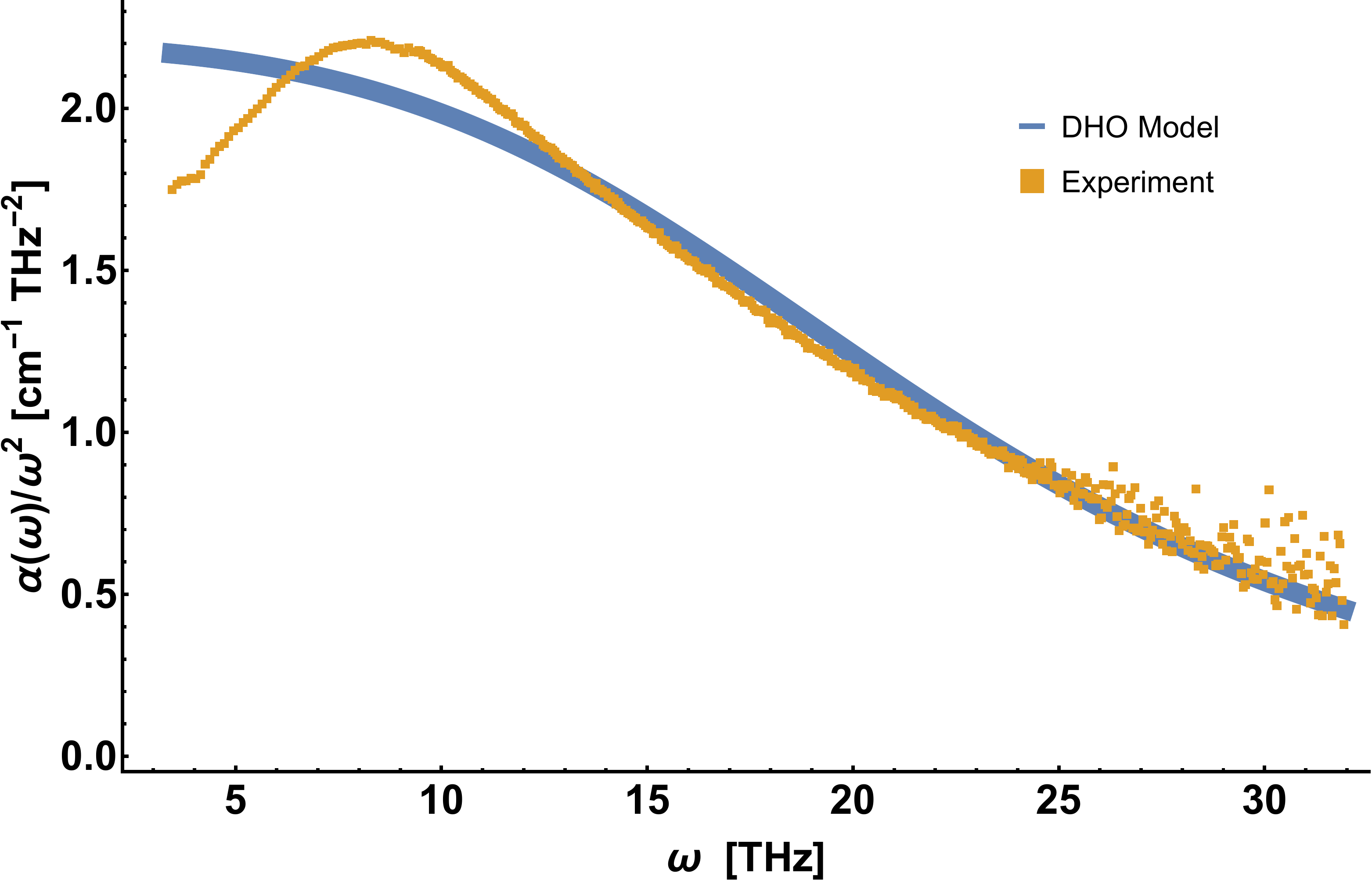}
\caption{Absorbance (normalized by Debye law) in the THz range. The yellow markers are the experimental data and the blue line is the fit with the constant damping model.}
\label{fig:visAbsor}
\end{figure}
As discussed in the main text, the damping mechanism is different depending on the frequency range we are looking at. At high energy (frequency/momentum), the microscopic details of the disorder are relevant, and the disorder-induced scattering is well approximated by a constant damping term $\Gamma(k)=\gamma$ as in the Drude model for electron conduction or in the Langevin equation for molecular motion in dense environment. In this regime, the damping is basically provided by the microscopic collisions in the localized motion of atoms. In Fig.\ref{fig:visAbsor} we show that this damping provides indeed a good approximation for the experimental data but only at large frequencies, much above the boson peak frequency $\omega_{BP}$. At low frequency, the experimental normalized absorbance turns down, while the DHO model with a constant damping cannot reproduce such trend.\\

\begin{figure}[h!]
 \centering
\includegraphics[width=0.9\linewidth]{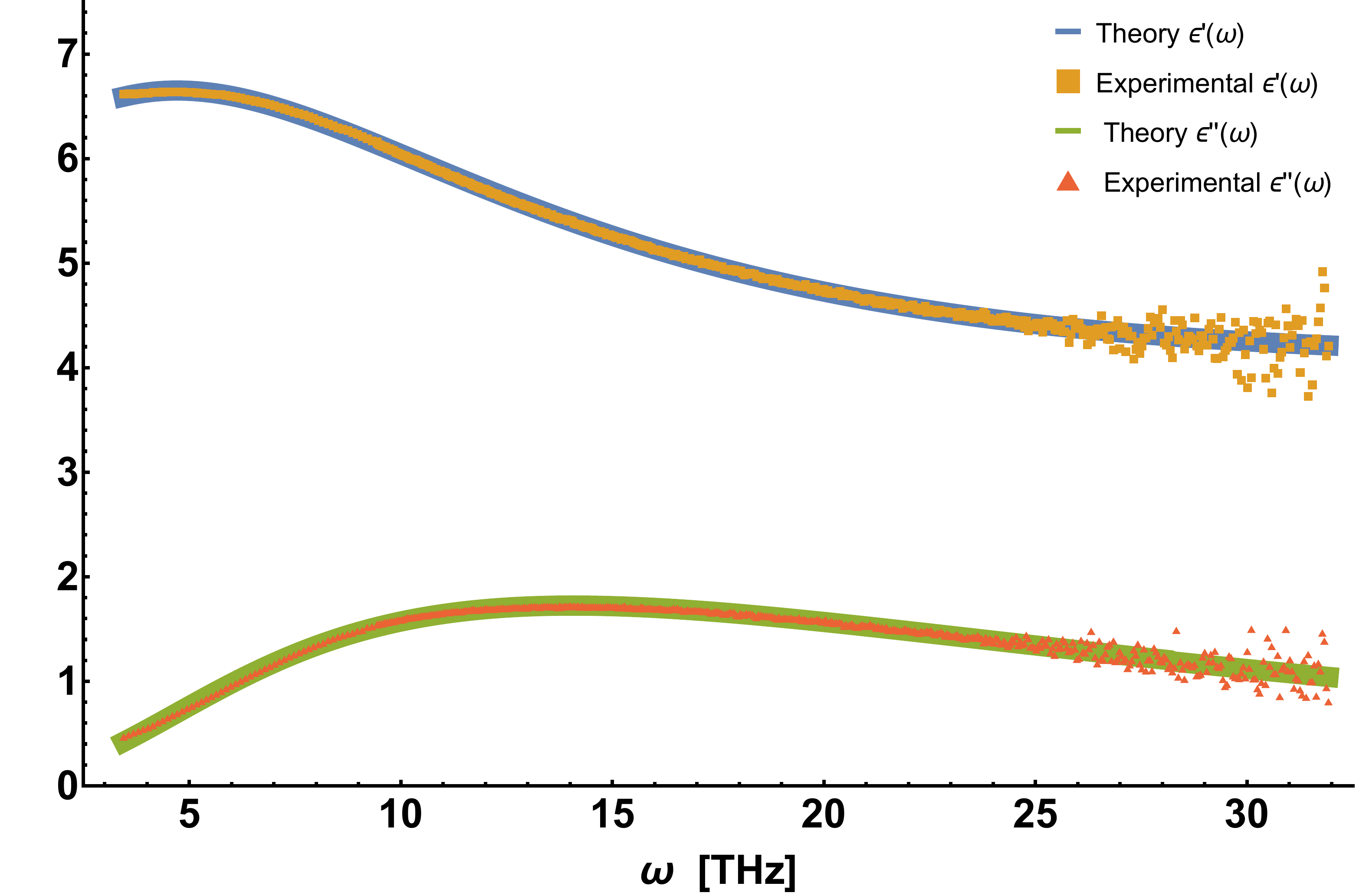}
\caption{Dielectric function of the soda lime glass in the THz range. The theoretical model uses the diffusive linewidth model,  Eq.\eqref{tutto} in the main text.}
\label{fig:visDiele}
\end{figure}

As explained in the main text, at low frequencies the nature of the linewidth can be well approximated by the  hydrodynamic expressions for diffusons:
\begin{equation}
    \Gamma(k)\,=\,D\,k^2.
\end{equation}
This mechanism comes from a coarse-grained description for which, on sufficiently large length scales, the effects of the microscopic scattering events are encoded in an effective ''diffusion'' dynamics. In order to have control over the full range of frequency, we build an interpolating model:
\begin{equation}
    \Gamma(k)\,=\,\dfrac{\gamma \,D\, k^{2}}{\gamma + D\, k^{2} }
\end{equation} 
which smoothly crosses over between the two, low-$k$ and high-$k$, regimes.
Using this model, we are able to accurately fit the experimental data across the whole range of frequencies. This is emphasized in Fig.\ref{fig:visDiele} where the full set of experimental data is shown and compared to our theory. 

\subsection*{Derivation of the Phonon-Polariton Dispersion Relations}
In order to derive our main relation Eq.\eqref{Eq. dispersion - damping function1}, we start by writing down the system of coupled dynamical equations for the relative (partially-)charged-particle displacement $\vec{u}$, the  polarization vector $\vec{P}$ and the EM fields $\vec{E},\vec{H}$.
\begin{equation}
    \label{eq: dispola&maxwell with damping function}
    \begin{cases}
        \Ddot{\Vec{u}}=b_{11}\Vec{u}- \Gamma\left(k\right) \dot{\Vec{u}}+b_{12}\Vec{E} \\
        \Vec{P}=b_{21}\Vec{u}+b_{22}\Vec{E} \\
        \nabla\cdot(\Vec{E}+4\pi\Vec{P})=0 \\
        \nabla\cdot\Vec{H}=0 \\
        \nabla \times \Vec{E} =  -\frac{1}{c}\dot{\Vec{H}}\\
        \nabla \times \Vec{H} = \frac{1}{c}( \dot{\Vec{E}}+4\pi\dot{\Vec{P}} )
    \end{cases}
\end{equation}
in which we importantly add an effective damping term $\Gamma(k)$ which encodes the effects of disorder and dissipation on the atomic motion. Going to Fourier space and identifying the coefficient $b_{11}$ with the characteristic mechanical oscillation frequency $\omega_0^2$ ~\cite{Born-Huang}, the equations can be written as
    \begin{equation}
    \label{ff}
    \begin{cases}
        \Vec{u}=-\dfrac{b_{12}}{\omega^{2}-\omega_{0}^{2}+i\omega \Gamma\left(k\right)}\Vec{E} \\
        \Vec{P}=\left(b_{22}-\dfrac{b_{12}b_{21}}{\omega^{2}-\omega_{0}^{2}+i\,\omega\, \Gamma\left(k\right)}\right)\Vec{E} \\
        \Vec{k}\cdot\Vec{E}\left(1+ 4\pi\, b_{22}-\dfrac{4\pi b_{12}b_{21}}{\omega^{2}-\omega_{0}^{2}+i\,\omega\, \Gamma\left(k\right)}\right)=0 \\
        \Vec{k}\cdot\Vec{H}=0 \\
        \Vec{k} \times \Vec{E} =  \frac{\omega}{c}\Vec{H}\\
        \Vec{k} \times \Vec{H} = -\frac{\omega}{c}( \Vec{E}+4\pi\Vec{P} )
    \end{cases}
\end{equation}
Using the known relation $\vec{D}=\vec{E}+4\pi\vec{\mathcal{P}}= \varepsilon(\omega)\vec{E}$, we can substitute the unknown parameters $b_{12},b_{21},b_{22}$ in terms of the dielectric constant. We denote:
\begin{equation}
    \varepsilon(\omega=0)\equiv \varepsilon_0\,,\quad \quad \varepsilon(\omega\rightarrow \infty)\equiv \varepsilon_\infty
\end{equation}
and using these definitions, the third equation in Eq.\eqref{ff} can be re-written as: 
 \begin{equation}
     \Vec{k}\cdot\Vec{E}\left(\varepsilon_{\infty} -\dfrac{\omega_{0}^{2}\left(\varepsilon_{0}-\varepsilon_{\infty}\right)}{\omega^{2}-\omega_{0}^{2} +i\,\omega\,\Gamma\left(k\right)}\right)=0. \label{gg}
 \end{equation}
At this point, a comment is in order. The scalar product $\Vec{k}\cdot\Vec{E}$ distinguishes between two different types of modes:
\begin{align}
    &\Vec{k}\cdot\Vec{E}\,\neq\,0\,\quad \quad \rightarrow \text{LO modes}\,,\\
    &\Vec{k}\cdot\Vec{E}\,=\,0\,\quad \quad \rightarrow \text{TO modes}\,.
\end{align}
Starting from the LO modes and assuming $\vec{k} \cdot \vec{E} \neq 0$, the Eq.\eqref{gg} implies
 \begin{equation}
 \begin{split}
     &\varepsilon_{\infty} =\dfrac{\omega_{0}^{2}\left(\varepsilon_{0}-\varepsilon_{\infty}\right)}{\omega^{2}-\omega_{0}^{2} +\,i\,\omega\,\Gamma\left(k\right)} \\
     &  \omega_{LO}^{2} +i\,\omega_{LO}\,\Gamma\left(k\right) -\omega_{0}^{2}\,\dfrac{\varepsilon_{0}}{\varepsilon_{\infty}} =0 \\
 \end{split}
 \end{equation}
 where we have indicated the frequency of the mode $\omega= \omega_{LO}$.\\
 
 Moving on to the TO modes and taking $\vec{k}$ and $\vec{E}$ orthogonal, we can observe from Eq.\eqref{ff} that the magnetic field $\vec{H}$ is orthogonal to both the vectors $\vec{k},\vec{E}$. As a consequence, the fifth and sixth equations in Eq.\eqref{ff} become a coupled relation between the amplitudes:
 
 \begin{equation*}
     \begin{cases}
        k\,E =  \dfrac{\omega_{TO}}{c}H\\
        k\,H = \dfrac{\omega_{TO}}{c}E\left(\varepsilon_{\infty} -\dfrac{\omega_{0}^{2}\left(\varepsilon_{0}-\varepsilon_{\infty}\right)}{\omega_{TO}^{2}-\omega_{0}^{2} +i\,\omega\,\Gamma\left(k\right)}\right)
     \end{cases}
     \end{equation*}
     \begin{equation}
     \Rightarrow \quad
     k^{2}c^{2} =  \omega_{TO}^{2}\left(\varepsilon_{\infty} -\dfrac{\omega_{0}^{2}\left(\varepsilon_{0}-\varepsilon_{\infty}\right)}{\omega_{TO}^{2}-\omega_{0}^{2} +i\,\omega\,\Gamma\left(k\right)}\right)
 \end{equation}

After simple algebraic manipulations, we finally obtain the quartic equation \eqref{Eq. dispersion - damping function1} presented in the main text:
\begin{align}
\label{Eq. dispersion - damping function}
  &  \omega^{4}\varepsilon_{\infty} + i \omega^{3} \Gamma\left(k\right) \varepsilon_{\infty} - \omega^{2}\left(\omega_{0}^{2}\varepsilon_{0} + k^{2}c^{2} \right) \nonumber\\&-i  \Gamma\left(k\right) k^{2}c^{2} \omega + \omega_{0}^{2} k^{2}c^{2} = 0 
\end{align}
where for simplicity we have omitted the label $TO$ which stands for ''transverse optical''.

\subsection{Effects of Disorder and Damping on the Phonon-Polariton Dispersion Relation}
\begin{figure}[htb]
    \centering
        \includegraphics[width=0.7\linewidth]{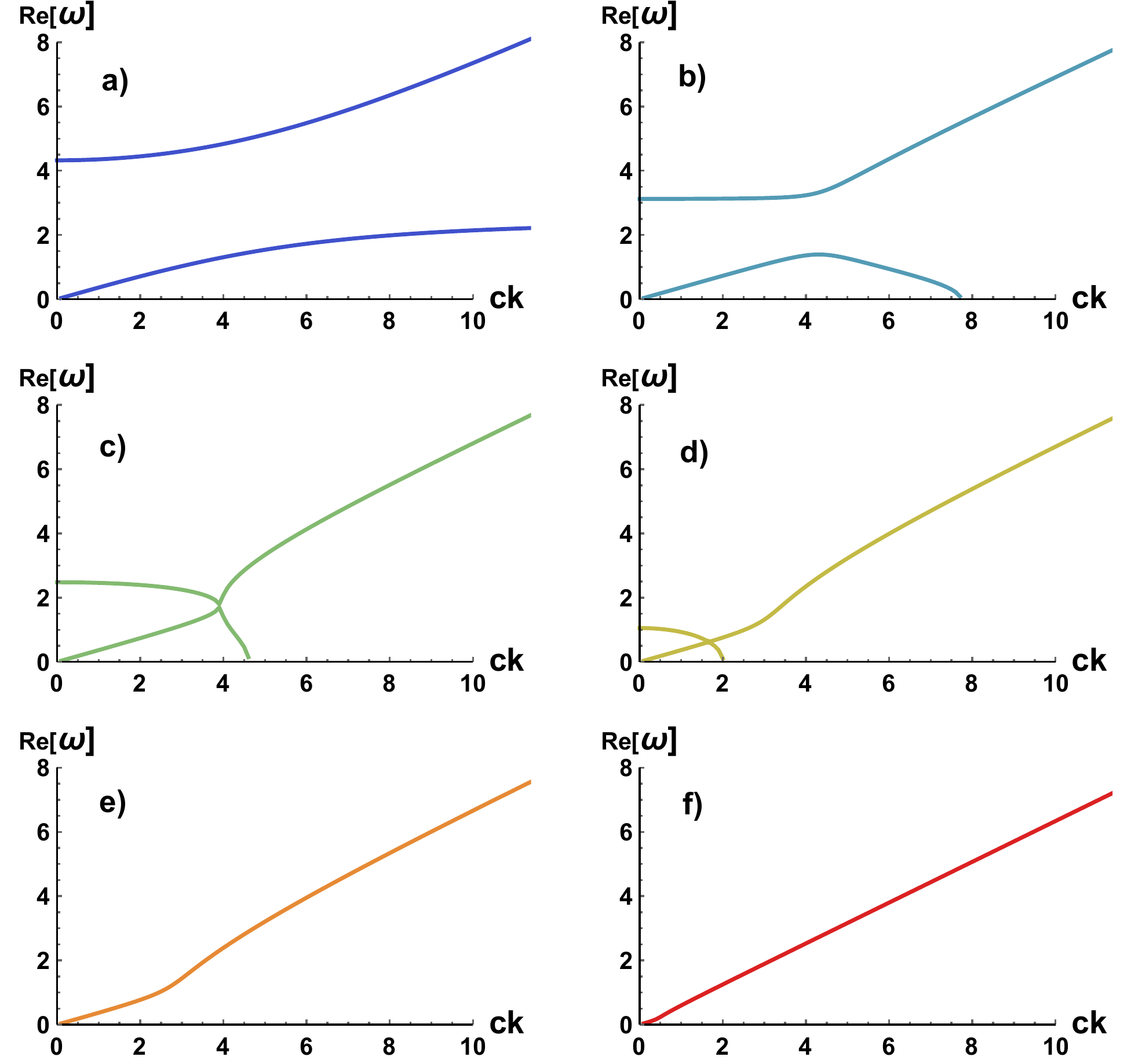}\\
        
        \vspace{0.5cm}
        \noindent\rule{3cm}{0.4pt}
        \vspace{0.5cm}
        
        \includegraphics[width=0.7\linewidth]{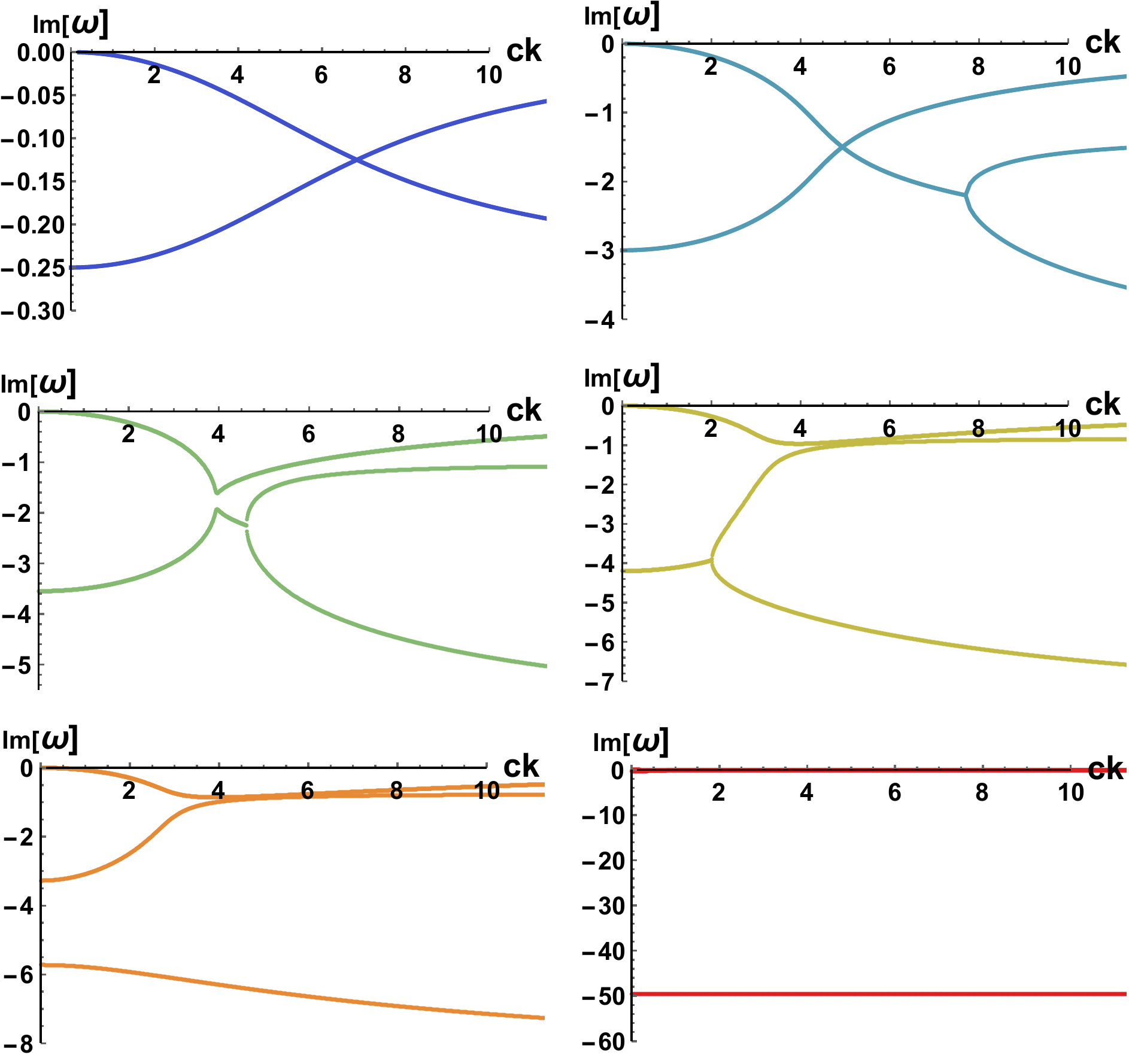}
    \caption{The dispersion relation of the excitations in the damped model by changing the damping parameter $\gamma =0.5,2.5,6,7.1,8.4,9,50$ from panel a) to panel f). \textbf{Top: }The real part $\mathrm{Re}(\omega)$ in function of the momentum $k$. \textbf{Bottom: }The imaginary part $\mathrm{Im}(\omega)$ in function of the momentum $k$.}
    \label{fig1}
\end{figure}

We start with the generic fourth-order equation which we derived in the previous section
\begin{align}
  &  \omega^{4}\varepsilon_{\infty} + i \omega^{3} \Gamma\left(k\right) \varepsilon_{\infty} - \omega^{2}\left(\omega_{0}^{2}\varepsilon_{0} + k^{2}c^{2} \right) \nonumber\\&-i  \Gamma\left(k\right) k^{2}c^{2} \omega + \omega_{0}^{2} k^{2}c^{2} = 0 
\end{align}
where the disorder and damping effects are effectively encoded in the momentum dependent parameter $\Gamma(k)$. Let us start by reminding the reader about the known results in absence of any damping mechanism, $\Gamma(k)=0$, which was derived in \cite{Huang-1950,Huang-1951}. In this simple case, the solution can be written concisely as:
\begin{equation}
    \omega\,=\,\sqrt{\frac{c^2 k^2\pm\sqrt{c^4 k^4+2 c^2 k^2 \omega_0^2 (\epsilon_0-2 \epsilon_\infty)+\epsilon_0^2 \omega_0^4}+\epsilon_0 \omega_0^2}{2\,\epsilon_\infty}}
\end{equation}
The two modes display the repulsion phenomenon which is typical of the polariton dynamics and is due to the electromagnetic interactions encoded in the non-trivial dieletric constant ($\epsilon_\infty \neq \epsilon_0$). This behaviour is very similar to the one is displayed in panel a) of Figure \ref{fig1} for a concrete choice of parameters with small damping. Obviously, this is an idealized situation in which all the effects which originate from internal scattering events are neglected.
\\
\begin{figure}[t]
     \centering
        \includegraphics[width=0.7\linewidth]{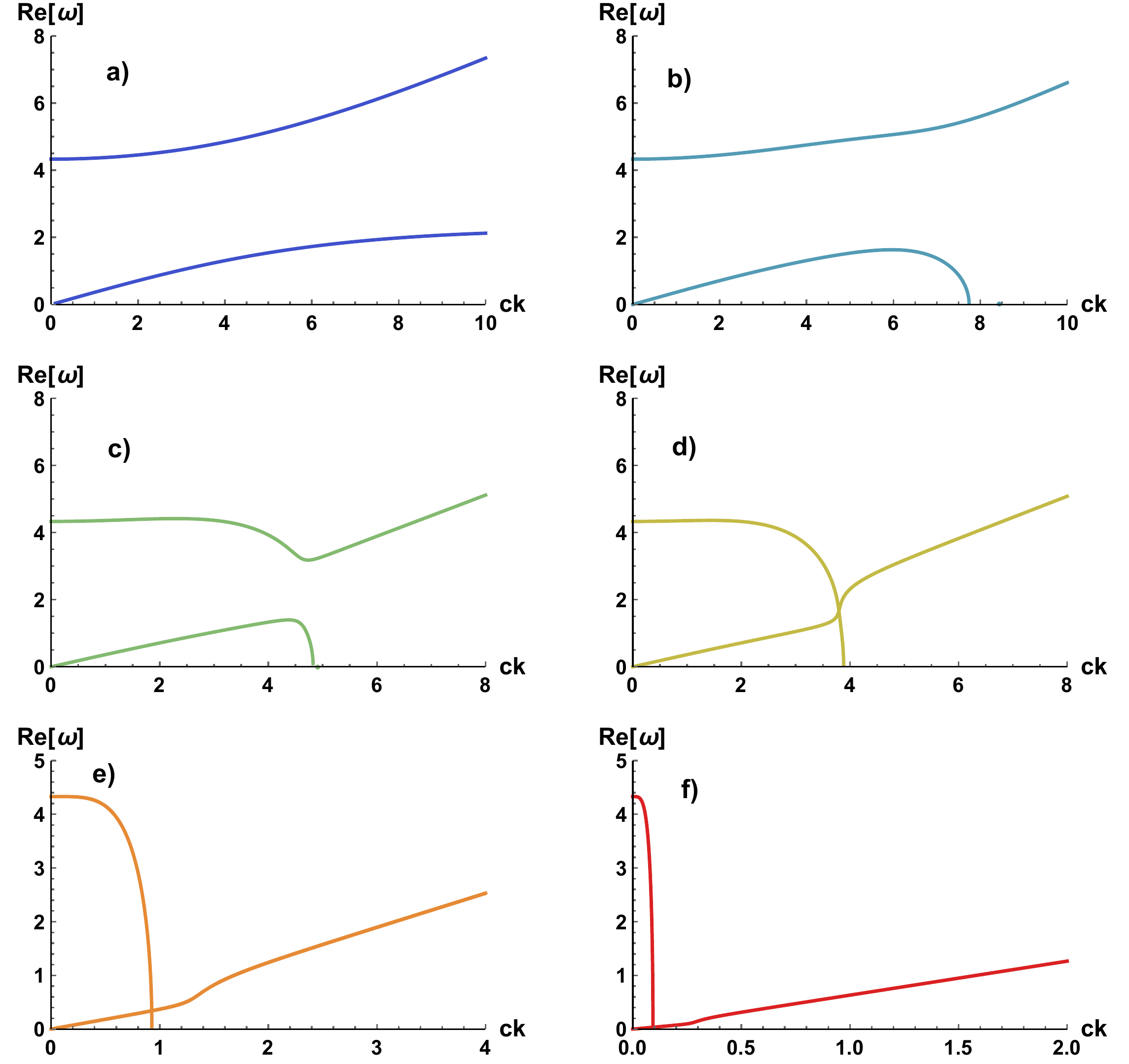}\\
        
        \vspace{0.5cm}
        \noindent\rule{3cm}{0.4pt}
        \vspace{0.5cm}
        
        \includegraphics[width=0.7\linewidth]{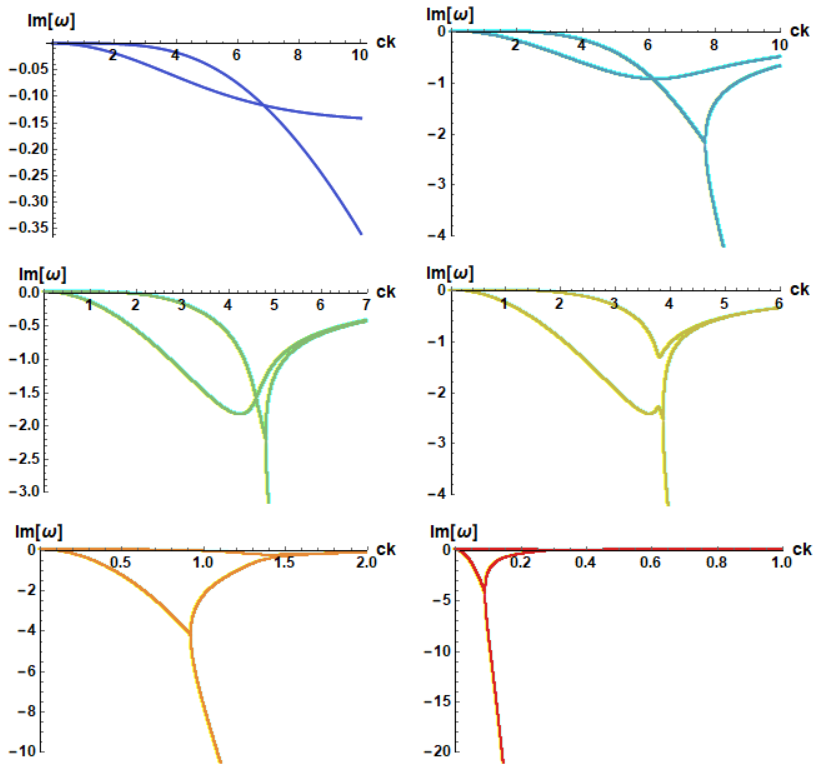}
    \caption{The dispersion relation of the excitations in the diffusive model by changing the diffusion constant $D=0.01,0.1,0.3,0.5,10,10^3$. \textbf{Top: }The real part $\mathrm{Re}(\omega)$ in function of the momentum $k$. \textbf{Bottom: }The imaginary part $\mathrm{Im}(\omega)$ in function of the momentum $k$.}
    \label{fig2}
\end{figure}

As a first step forward, let us consider the situation in which the optical phonons have a finite and momentum independent relaxation time:
\begin{equation}
    \tau^{-1}\,=\,\Gamma(k=0)=\gamma
\end{equation}
which determines their lifetime and mean free path. Here, we take an effective field theory perspective and we do not discuss the microscopic origin of this relaxation time. Several are the physical mechanisms that can contribute to this effect. Theoretically, this relaxation time implies the non-conservation of momentum, which now dissipates at a rate $\gamma$, exactly as in the simple Drude model for electric conduction \cite{kittel2004introduction} or in the Langevin equation for Brownian motion in liquids. This relaxation time approximation can be formally derived using Boltzmann equation and kinetic theory \cite{soto2016kinetic} and it is valid only in the regime when $\tau$ is large enough. The dynamics of the low energy modes is displayed in Fig. \ref{fig1} upon increasing the relaxation rate $\gamma \in [0,50]$ from panel a) to panel f). For small $\gamma \ll \omega_0$, the gapless mode acquires a small damping $\mathrm{Im}(\omega)(k=0) \neq 0$ which grows with $\gamma$. This mode is not anymore a hydrodynamic mode. The other gapped mode does not acquire a finite damping and remains diffusive at low momentum. When the damping parameter becomes comparable with the characteristic frequency of the gapped mode $\gamma \sim \omega_0$, the two modes attract each other and they move closer as shown in panel b) of Fig.\ref{fig1}. When $\gamma \geq 2 \omega_0$, the dynamics is not anymore \textit{under-damped} and the modes merge producing a complicated pattern shown in the panels c) and d) of Fig.\ref{fig1}. Finally, in the limit $\gamma \gg \omega_0$ (\textit{over-damped} regime), the sound mode gets completely destroyed and it acquires a very large damping. Its lifetime becomes very short and it completely disappears from the dynamics (see panel e) in Fig.\ref{fig1}). As a consequence, the ''photon root'' does not feel anymore its presence and the dispersion relation of the left mode goes back to the free case $\omega=\pm c k$, in which interactions are absent. This last step is shown in panel f) of Fig.\ref{fig1}.\\
Let us consider now a second and different case which will be more relevant for our discussion. More precisely, let us assume that the imaginary part of the vibrational mode is purely diffusive:
\begin{equation}
    \Gamma(k)\,=\,D\,k^2
\end{equation}
The parameter $D$ is the diffusion constant of the diffusons and is determined by elastic scattering events due to the disorder. Importantly, this choice is different with respect to the previous one in several aspects. The presence of diffusion does not imply the non-conservation of momentum~\cite{Ruocco2007}, nor the explicit breaking of any symmetry. The quasiparticle nature of the optical phonons gets lost across a ballistic to diffusive crossover (Ioffe-Regel crossover) \cite{ioffe1960non}. The phenomenon is shown in Fig.\ref{example}.\\
\begin{figure}[b!]
    \centering
    \includegraphics[width=0.8\linewidth]{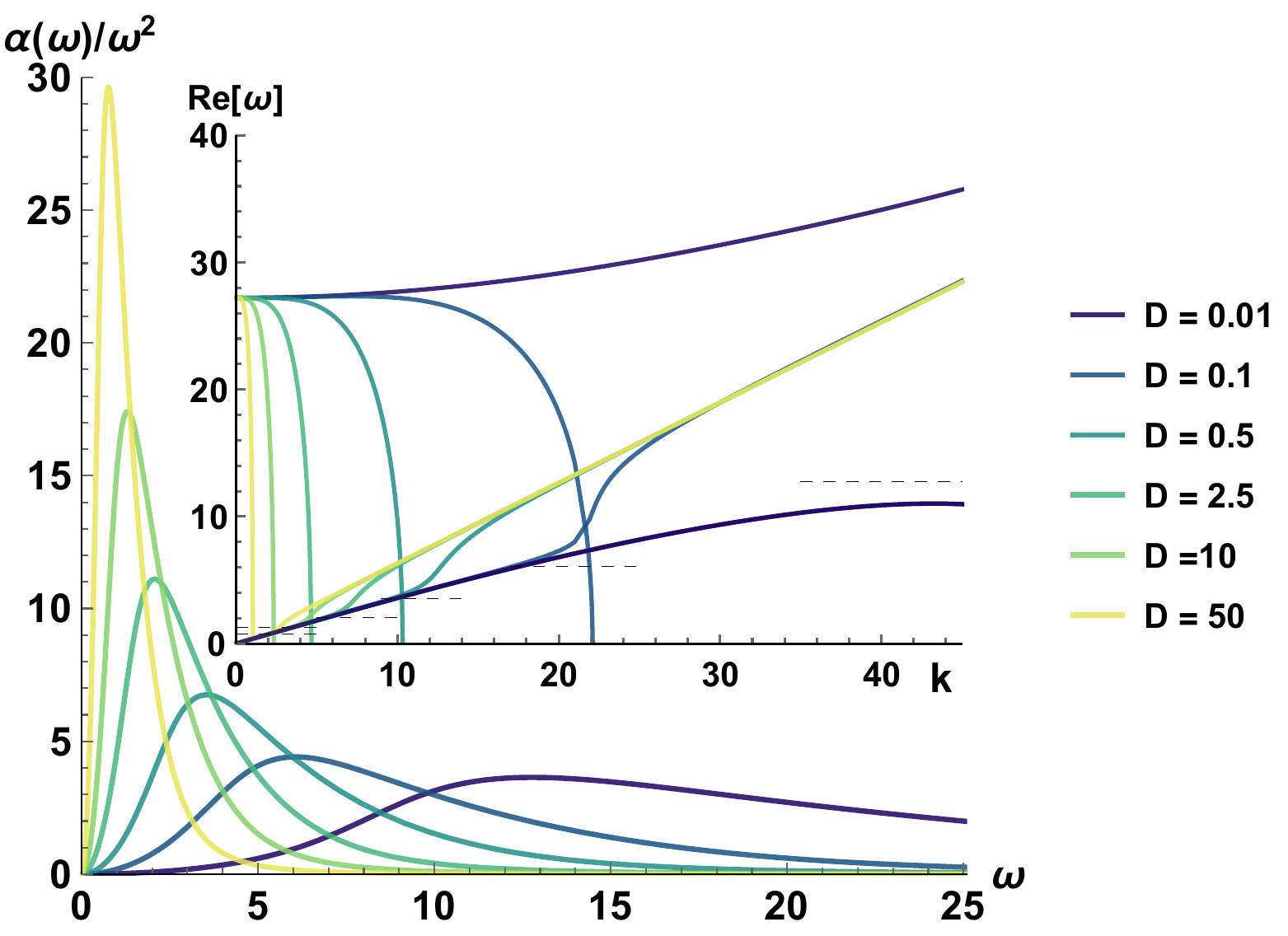}
    \caption{The Debye normalized absorbance coefficient in function of the frequency $\omega$ assuming a diffusive damping $\Gamma(k)=D k^2$. We vary the diffusion constant $D$ as indicated in the legend. We fix $\varepsilon_\infty=2.5$, $\varepsilon_0=7.5$, $\Omega=5 \pi$. The inset shows the corresponding dispersion relation of the phonon-polariton modes where the dashed horizontal lines indicate the position of the maxima in the absorbance -- the BP frequency $\omega_{BP}$.}
    \label{fig10}
\end{figure}
The dynamics of the real part of the modes is very similar to the previous case (compare the top panels of Fig.\ref{fig1} and Fig.\ref{fig2}). The difference is nevertheless evident in the imaginary part of the modes. First, as already announced, both the modes (real and imaginary part) remain hydrodynamic, in the sense that both the imaginary parts vanish at zero momentum. Second, the difference is evident also comparing the situation at large damping $\gamma \gg 1$ with that at large diffusion $D\gg 1$. In the first case, one of the two modes disappear from the low energy dynamics because it becomes overdamped, with $\mathrm{Im}(\omega)\,\sim\,-\,\gamma$ very large. In the second case, at large diffusion constant, the two modes also stop to interact but this second mode becomes now totally diffusive $\omega\,\sim\,-\,i\,D\,k^2$ and therefore still present in the low energy dynamics of the system. This second hydrodynamic mechanism is crucial in our discussion since the effects of diffusion will be fundamental to give a complete theoretical understanding of the experimental data.\\

Finally, in Fig.\ref{fig10}, we show the Debye normalized absorbance predicted by theory using the diffusive model. The position of the BP moves towards lower frequency by increasing the diffusion constant $D$. This dynamics is consistent with the correlation of the BP frequency with the Ioffe-Regel crossover. In the inset, we show also the dispersion relation of the phonon-polariton to emphasize that the BP frequency and the frequency of band flattening $\omega_{flat}$ defined in the main text do not coincide.

\end{document}